\documentclass[a4paper,11pt,]{article}

\usepackage[left=2.5cm,
            right=2.5cm,
            top=2.5cm,
            bottom=2.5cm]{geometry}

\usepackage[T1]{fontenc}
\usepackage[utf8]{inputenc}
\usepackage[english]{babel}
\usepackage{amsmath,amsfonts,amssymb}
\usepackage{lmodern}
\usepackage[scaled]{helvet}
\usepackage[T1]{fontenc}
\usepackage[scaled=0.95]{inconsolata}

\usepackage[dvipsnames]{xcolor}
\definecolor{black50}{gray}{0.5} 
\definecolor{color0}{RGB}{0,0,0} 
\definecolor{color1}{RGB}{59,90,198} 
\definecolor{pnasbluetext}{RGB}{0,101,165}
\definecolor{pnasblueback}{RGB}{205,217,235}

\definecolor{macblue}{HTML}{185FAF}  
\definecolor{pnasbluetext}{RGB}{0,0,0} %

\usepackage{afterpage}

\usepackage{fontawesome5}  

\usepackage{afterpage}
\usepackage{ifpdf,ifxetex}
\ifpdf\else
  \ifxetex\else
    
    \pdfpagewidth=\paperwidth
    \pdfpageheight=\paperheight
\fi\fi
\usepackage{tikz}
\usepackage[framemethod=tikz]{mdframed}

\IfFileExists{flushend.sty}
    {\usepackage{flushend}}
    {\typeout{}
    \typeout{Package widetext error: Install the flushend package which is
    a part of sttools bundle. Available from CTAN.}
    \typeout{}
    \stop
    }

\IfFileExists{cuted.sty}
    {\usepackage{cuted}}
    {\typeout{}
    \typeout{Package widetext error: Install the cuted package which is
    a part of sttools bundle. Available from CTAN.}
    \typeout{}
    \stop
    }

{%
    \par
    \hfill\rule[-6pt]{0.4pt}{6.4pt}%
    \rule{\dimexpr(0.5\textwidth-0.5\columnsep-1pt)}{0.4pt}
    \end{strip}
}

\usepackage[colorlinks=true, allcolors=pnasbluetext]{hyperref}

\newcommand{\footerfont}{\normalfont\sffamily\footnotesize}
\newcommand{\titlefont}{\fontfamily{lmss}\bfseries\LARGE}

\newcommand{\addinfofont}{\normalfont\sffamily\fontsize{7}{8}\selectfont}
\newcommand{\absfont}{\normalfont\sffamily\small}
\newcommand{\keywordsfont}{\normalfont\sffamily\small}

\usepackage{xifthen}
\newboolean{shortarticle}
\newboolean{singlecolumn}

\def\useignorespacesandallpars#1\ignorespaces\fi{%
  #1\fi\ignorespacesandallpars}
\def\ignorespacesandallpars{%
  \@ifnextchar\par
  {\expandafter\ignorespacesandallpars\@gobble}%
  {}%
}

\newboolean{displaycopyright}
\setboolean{displaycopyright}{false} 
\usepackage{textcomp} 

\usepackage{graphicx,xcolor}
\graphicspath{{Figures/}}
\usepackage{colortbl}
\usepackage{booktabs}
\usepackage[noend]{algpseudocode}
\usepackage{changepage}

\usepackage[labelfont={bf,sf}, labelsep=period]{caption}

\usepackage{natbib}
\DeclareCaptionFormat{pnasformat}{\normalfont\sffamily\small #1#2#3}
\usepackage{etoolbox}
\AtBeginEnvironment{tabular}{\small}

\usepackage{algorithm}
\usepackage[noend]{algpseudocode}
\algnewcommand\algorithmicinput{\textbf{Input:}}
\algnewcommand\algorithmicoutput{\textbf{Output:}}
\algnewcommand\Input{\item[\algorithmicinput]}%
\algnewcommand\Output{\item[\algorithmicoutput]}%
\algnewcommand\Step{\item[]}%

\usepackage{fancyhdr}  
\usepackage{lastpage}  
\def\footercontents#1{\def\@footercontents{#1}}
\footercontents{Package Vignette}
\newcommand{\printfootercontents}{\@footercontents}

\makeatletter
\fancypagestyle{firststyle}{
   \fancyfoot[R]{\footerfont \printfootercontents\hspace{7pt}|\hspace{7pt}\textbf{\@ifundefined{@documentdate}{\today}{\@documentdate}}\hspace{7pt}|\hspace{7pt}\textbf{\thepage\textendash\pageref{LastPage}}}
   \fancyfoot[L]{\footerfont\@ifundefined{@doifooter}{}{\@doifooter}}
}
\makeatother

\makeatletter
\appto{\titlefont}{\color{macblue}}

\usepackage[explicit,nobottomtitles]{titlesec}
\titleformat{\section}
  {\large\sffamily\bfseries\color{macblue}}
  {\thesection.}
  {0.5em}
  {#1}
  []
\titleformat{name=\section,numberless}
  {\large\sffamily\bfseries\color{macblue}}
  {}
  {0em}
  {#1}
  []
\titleformat{\subsection}
  {\sffamily\bfseries\color{macblue}} 
  {\thesubsection.}
  {0.5em}
  {#1}
  []

\titleformat{\subsubsection}[runin]
  {\sffamily\bfseries\color{macblue}}
  {\thesubsubsection.}
  {0.5em}
  {#1 \quad }
  []

\titleformat{\paragraph}[runin]
  {\sffamily\bfseries}
  {}
  {0em}
  {#1 }

\titlespacing*{\section}{0pc}{3ex \@plus4pt \@minus3pt}{10pt}
\titlespacing*{\subsection}{0pc}{2.5ex \@plus3pt \@minus2pt}{5pt}
\titlespacing*{\subsubsection}{0pc}{2ex \@plus2.5pt \@minus1.5pt}{2pt}
\titlespacing*{\paragraph}{0pc}{1.5ex \@plus2pt \@minus1pt}{12pt}

\newcommand{\additionalelement}[1]{\def\@additionalelement{#1}}
\newcommand{\addinfo}[1]{\def\@addinfo{#1}}
\newcommand{\doifooter}[1]{\def\@doifooter{#1}}
\newcommand{\documentdate}[1]{\def\@documentdate{#1}}
\newcommand{\leadauthor}[1]{\def\@leadauthor{#1}}
\newcommand{\etal}[1]{\def\@etal{#1}}
\newcommand{\keywords}[1]{\def\@keywords{#1}}
\newcommand{\authorcontributions}[1]{\def\@authorcontributions{#1}}
\newcommand{\authordeclaration}[1]{\def\@authordeclaration{#1}}
\newcommand{\equalauthors}[1]{\def\@equalauthors{#1}}
\newcommand{\correspondingauthor}[1]{\def\@correspondingauthor{#1}}
\newcommand{\significancestatement}[1]{\def\@significancestatement{#1}}
\newcommand{\matmethods}[1]{\def\@matmethods{#1}}
\newcommand{\acknow}[1]{\def\@acknow{#1}}

\def\xabstract{abstract}
\long\def\abstract#1\end#2{\def\two{#2}\ifx\two\xabstract
\long\gdef\theabstract{\ignorespaces#1}
\def\go{\end{abstract}}\else
\typeout{^^J^^J PLEASE DO NOT USE ANY \string\begin\space \string\end^^J
COMMANDS WITHIN ABSTRACT^^J^^J}#1\end{#2}
\gdef\theabstract{\vskip12pt BADLY FORMED ABSTRACT: PLEASE DO
NOT USE {\tt\string\begin...\string\end} COMMANDS WITHIN
THE ABSTRACT\vskip12pt}\let\go\relax\fi
\go}

\newcommand{\abscontent}{
\noindent
\parbox{\dimexpr\linewidth}{%
  \vskip3pt\absfont%
  {\bfseries Abstract} \theabstract
}%
\vskip10pt%
\noindent
\parbox{\dimexpr\linewidth}{%
{
 \keywordsfont
 \@ifundefined{@keywords}{}{{\bfseries Keywords} \@keywords}}%
}
\vskip12pt%
}

\newcommand{\abscontentformatted}{
\abscontent
}

\renewcommand{\@maketitle}{%
\def\And{\vskip5pt}
{%
\ifthenelse{\boolean{shortarticle}}
  {\ifthenelse{\boolean{singlecolumn}}{}{
    {\raggedright\baselineskip= 24pt\titlefont \@title\par}%
    \vskip10pt
    {\raggedright \@author\par}
    \vskip8pt
    {\raggedright \addinfofont \@ifundefined{@addinfo}{}{\@addinfo}\par}
    \vskip12pt%
    }}
  {
    \vskip10pt%
    {\raggedright\baselineskip= 24pt\titlefont \@title\par}%
    \vskip10pt
    {\raggedright \@author\par}
    \vskip8pt
    {\raggedright \addinfofont \@ifundefined{@addinfo}{}{\@addinfo}\par}
    \vskip12pt
    {%
    \abscontent
    }%
    \vskip25pt%
  }%
}%
}

\usepackage[flushmargin,ragged]{footmisc}
\renewcommand{\footnoterule}{
  \kern -3pt
  {\color{black50} \hrule width 72pt height 0.25pt}
  \kern 2.5pt
}

\titleclass{\acknow@section}{straight}[\part]
\newcounter{acknow@section}
\providecommand*{\toclevel@acknow@section}{0}
\titleformat{\acknow@section}[runin]
   {\sffamily\normalsize\bfseries\color{macblue}}
   {}
   {0em}
   {#1.}
   []
\titlespacing{\acknow@section}
	{0pt}
	{3.25ex plus 1ex minus .2ex}
	{1.5ex plus .2ex}

\newcommand{\showacknow}{
\@ifundefined{@acknow}{}{
\vskip 3.25ex plus 1ex minus .2ex
\noindent{\sffamily\normalsize\bfseries Acknowledgments.\hspace{1.5ex plus .2ex}}
\small\@acknow}
}

\titleclass{\matmethods@section}{straight}[\part]
\newcounter{matmethods@section}
\providecommand*{\toclevel@matmethods@section}{0}
\titleformat{\matmethods@section}
   {\sffamily\normalsize\bfseries}
   {}
   {0em}
   {#1}
   []
\titlespacing{\matmethods@section}
	{0pt}
	{3.25ex plus 1ex minus .2ex}
	{1.5ex plus .2ex}
\newcommand{\showmatmethods}{
\@ifundefined{@matmethods}{}{\matmethods@section{Materials and Methods}{\small\noindent\@matmethods}}
}

\usepackage{enumitem} 

\makeatletter
\DeclareRobustCommand\code{\bgroup\@noligs\@codex}
\def\@codex#1{\texorpdfstring%
{{\normalfont\ttfamily\hyphenchar\font=-1 #1}}%
{#1}\egroup}
\makeatother

\usepackage{fancyvrb}

\DefineVerbatimEnvironment{Highlighting}{Verbatim}{commandchars=\\\{\}}

\definecolor{codeboxbg}{RGB}{246,246,246}
\global\mdfdefinestyle{codebox}{%
  linewidth=0pt,
  backgroundcolor=codeboxbg,
  innerleftmargin=2pt,
  innerrightmargin=2pt}

\global\mdfdefinestyle{resultbox}{%
  linewidth=0pt,
  backgroundcolor=codeboxbg,
  innerleftmargin=2pt,
  innerrightmargin=2pt}

\setboolean{shortarticle}{true}

\usepackage{float}
\floatstyle{plain}
\newfloat{sigstatement}{b!}{sst}

\additionalelement{%
\afterpage{\begin{sigstatement}
\sffamily
\mdfdefinestyle{pnassigstyle}{linewidth=0.7pt,backgroundcolor=pnasblueback,linecolor=pnasbluetext,fontcolor=pnasbluetext,innertopmargin=6pt,innerrightmargin=6pt,innerbottommargin=6pt,innerleftmargin=6pt}
\@ifundefined{@significancestatement}{}{%
	\begin{mdframed}[style=pnassigstyle]%
	\section*{Significance Statement}%
	\@significancestatement
	\end{mdframed}}
\scriptsize
\@ifundefined{@authorcontributions}{}{\@authorcontributions}
\vskip5pt%
\@ifundefined{@authordeclaration}{}{\@authordeclaration}
\vskip5pt%
\@ifundefined{@equalauthors}{}{\@equalauthors}
\vskip5pt%
\@ifundefined{@correspondingauthor}{}{\@correspondingauthor}
\end{sigstatement}}
}

\usepackage[bitstream-charter]{mathdesign} 

\usepackage{upquote} 

\usepackage{longtable}

\footercontents{\textsf{\bfseries ar\textcolor[HTML]{B31B1B}{X}iv}}

\newcommand{\Info}{\boldsymbol{I}}

\newcommand{\x}{\boldsymbol{x}}
\newcommand{\y}{\boldsymbol{y}}

\newcommand{\betab}{\boldsymbol{\beta}}

\renewcommand{\S}{\boldsymbol{S}}
\newcommand{\T}{{}^{\!\top}}
\renewcommand{\hat}[1]{\widehat{#1}}

\DeclareMathOperator{\logit}{logit}
\DeclareMathOperator{\logistic}{logistic}
\newcommand{\Beta}{\text{Beta}}
\DeclareMathOperator{\Exp}{\mathbb{E}}
\DeclareMathOperator{\Var}{\mathbb{V}}
\newcommand{\covindex}{\text{COVINDEX}}
\usepackage{booktabs}
\usepackage{longtable}
\usepackage{array}
\usepackage{multirow}
\usepackage{wrapfig}
\usepackage{float}
\usepackage{colortbl}
\usepackage{pdflscape}
\usepackage{tabu}
\usepackage{threeparttable}
\usepackage{threeparttablex}
\usepackage[normalem]{ulem}
\usepackage{makecell}
\usepackage{xcolor}

\begin{document}

\title{A COVINDEX based on a GAM beta regression model with an application to the COVID-19 pandemic in Italy}

\author{
  \vspace*{2ex}
    \small\textbf{Luca Scrucca}
   \\
  \footnotesize
    Dipartimento di Economia \\
    Università degli Studi di Perugia \\
  Via A. Pascoli 20, 06123 Perugia, Italy \\
  \textcolor[HTML]{005392}{\faEnvelope}\;\;\href{mailto:luca.scrucca@unipg.it}{\nolinkurl{luca.scrucca@unipg.it}} \\
  \textcolor[HTML]{A6CE39}{\faOrcid}\;\;\url{https://orcid.org/0000-0003-3826-0484}\\
  }

\begin{abstract}
Detecting changes in COVID-19 disease transmission over time is a key indicator of epidemic growth.
Near real-time monitoring of the pandemic growth is crucial for policy makers and public health officials who need to make informed decisions about whether to enforce lockdowns or allow certain activities.
The effective reproduction number \(R_t\) is the standard index used in many countries for this goal. However, it is known that due to the delays between infection and case registration, its use for decision making is somewhat limited.
In this paper a near real-time COVINDEX is proposed for monitoring the evolution of the pandemic. The index is computed from predictions obtained from a GAM beta regression for modelling the test positive rate as a function of time. The proposal is illustrated using data on COVID-19 pandemic in Italy and compared with \(R_t\). A simple chart is also proposed for monitoring local and national outbreaks by policy makers and public health officials.
\end{abstract}

\keywords{pandemic surveillance; GAM beta regression; COVINDEX; public-health decision-making}

\maketitle
\thispagestyle{firststyle}
\ifthenelse{\boolean{shortarticle}}{\ifthenelse{\boolean{singlecolumn}}{\abscontentformatted}{\abscontent}}{}


\section{Introduction}
\label{introduction}

The World Health Organization (WHO) declared coronavirus disease (COVID-19) a pandemic on 11 March 2020.
Since then, most countries around the world have addressed this threat by implementing various strategies to fight the pandemic.
From simple preventive measures, such as case identification and contact tracing, quarantine and isolation, to more severe strategies based on general lockdowns of all non-essential economical and social activities.
Since public health decision-making requires the balancing of numerous, and often conflicting, factors, a timely and data-informed decision making process appears crucial.

Several studies have been recently devoted to the analysis of COVID-19 data. 
Referring to the Italian situation, \citet{Sebastiani:Massa:Riboli:2020} evaluated the impact of government measures on the evolution of pandemic. 
\citet{Girardi:etal:2020} used robust dose-response curves to predict the contagion dynamics of COVID-19, while \citet{AlaimoDiLoro:etal:2021} proposed an extended Generalized Linear Model based on the Richards' curve to model and predict incidence indicators. 
A Poisson autoregressive model was discussed by \citet{Agosto:etal:2021} to monitor the time evolution of the COVID-19 contagion curve, while \citet{Bartolucci:Farcomeni:2021} introduced a spatio-temporal model based on discrete latent variables for the analysis of weekly positive rates. 
Finally, \citet{Farcomeni:etal:2021} investigated an ensemble approach for short-term prediction of occupancy of intensive care units due to COVID-19 outbreak. 

The basic reproduction number, $R_0$, is an indicator of the epidemic's virulence.
It is defined as the average number of infections caused by an infected person when the whole population is susceptible, and for SARS-CoV-2 is between 2 and 3 \citep{Li:etal:2020, Hilton:Keeling:2020}.
As the pandemic evolves, the effective reproduction number $R_t$ is a more useful measure. This is the average number of infections that an infected person will cause. An $R_t$ above 1.0 indicates that the outbreak is growing, and below 1.0 means that it is shrinking. 
As a simple understood measure, $R_t$ is regularly published and discussed by the media, and it has been used in many countries, including Italy, to decide whether to tighten or loosen control measures.
However, $R_t$ suffers from several drawbacks when used to monitor the transmission of the disease over time, the main one being the delay with which it signals the evolution of the pandemic \citep{Gostic:etal:2020, Adam:2020}. Therefore, with a delay on the estimate of $R_t$ between ten days to two weeks, the use of $R_t$ as a near real-time decision-making tool appears rather pointless.
For further discussion on the risks caused by the misuse of the reproduction number in the COVID‐19 surveillance see \citet{Maruotti:Ciccozzi:Divino:2021}.

This paper introduces a COVID-19 index, called COVINDEX, which tries to assess whether the epidemic is growing, shrinking, or holding steady.
The proposed index is estimated by modelling the test positive rate (TPR) with a GAM beta regression model. TPR is an easily computed statistic, defined as the fraction of all COVID-19 tests performed on a given day that are actually positive. This metric can be used to understand the spread of the virus, but it also offers a measure of how adequately a country is testing.
TPR can be high if the number of positive tests is too high, but also if the number of total tests is too low. Most developed countries faced limited testing capacity during the initial phase of the pandemic, which resulted in high TPR values due to testing conducted primarily on symptomatic individuals. In the following months the ability to administer tests using PCR (polymerase chain reaction) or molecular swabs largely increased, leading to a situation that allows both symptomatic and asymptomatic individuals to be tested.
Although TPR can't be used for estimating incidence of the virus in the general population, a fundamental epidemic parameter that would require a carefully designed sampling plan, it can be used for monitoring the evolution of infection and transmission in the community. Higher positive rates suggest the need for further restrictions, such as wearing masks and physical distancing, to slow down the spread of the disease.
As a rule of thumb, World Health Organization recommended 5\% as the threshold for the percent positive rate to declare the COVID-19 transmission under control.

The main advantages of the proposed COVINDEX is the use of data routinely collected and its timely estimation which provides a near real-time tool to assess the effectiveness of interventions and to inform policy. Furthermore, since it is based on a statistical model, the associated uncertainty can be estimated.

The paper is organized as follows.
Section \ref{statistical-model-for-the-test-positive-rate} introduces the GAM beta regression model, its estimation and uncertainty assessment.
Section \ref{covindex-as-a-monitoring-and-decision-making-tool} describes the proposed COVINDEX and its usage for monitoring the pandemic evolution.
Section \ref{application-to-italian-covid-19-pandemic} includes a detailed analysis of COVID-19 pandemic in Italy, including the estimation of COVINDEX, from early March 2020 to the end of March 2021.
Section \ref{a-comparison-of-covindex-with-the-effective-reproduction-number} contains a comparison between the proposed COVINDEX and the effective reproduction number, showing the advantages of COVINDEX as near real-time monitoring tool.
The final section provides some concluding remarks.

\section{Statistical Model for the Test Positive Rate}
\label{statistical-model-for-the-test-positive-rate}

\subsection{GAM beta regression}
\label{gam-beta-regression}

Let $y_t$ be the test positive rate (TPR), defined as the ratio of the number of new positive cases $P_t$ to the number of tests $T_t$ at time $t$, with $t$ assuming integer values between $1$ and $n$, respectively, for the first and last day of the analyzed period. 
As a proportion TPR is naturally limited in the range $[0,1]$. Several approaches and models can be used for response variables that are expressed as proportions \citep{Douma:Weedon:2019}, and perhaps the most popular statistical model is the beta regression model \citep{Ferrari:CribariNeto:2004,Zeileis:CribariNeto:2010}.

Assume that TPR can be modelled by a beta distribution written as
$$
y_t \sim \Beta(\mu_t, \phi),
$$
with mean and variance of the beta distribution given, respectively, by
\begin{align*}
\Exp[y_t] & = \mu_t, \\
\intertext{and}
\Var[y_t] & = \frac{\mu_t(1-\mu_t)}{1+\phi}.
\end{align*}

Strictly speaking, the beta distribution can only model data in the open set $(0,1)$. If extreme values 0 and 1 can actually be observed, the inflated zero- and/or one beta distribution of \citet{Ospina:Ferrari:2010} could be used. 
Since in practice TPR rarely assumes a value of 0 and almost never 1, if needed, the simple approach proposed by \citet{Smithson:Verkuilen:2006} can be adopted by applying the data transformation $(y_t (n-1) + 0.5)/n$. The latter is the approach followed in this paper.

The mean $\mu_t$ can be expressed as a function of the linear predictor $\eta_t = \betab\T\x_t$, where $\betab$ is a $(p+1)$-dimensional vector of unknown regression coefficients (including the intercept), and $\x_t$ is the vector of observed values on $p$ predictors plus a one for the intercept.
Usually, the logistic function is used in beta regression, so we can write
$$
\mu_t = \logistic(\eta_t) = \frac {\exp(\eta_t)}{1+\exp(\eta_t)} = \frac{1}{1+\exp(-\eta_t)}.
$$

The inverse of the logistic function is the logit function, the so-called \emph{link} function in GLM terminology \citep{McCullaghNelder:1989}, given by:
$$
\logit(\mu_t) = \log\left(\frac{\mu_t}{1-\mu_t}\right) = \eta_t.
$$
 
Generalized Additive Models \citep[GAMs; ][]{Hastie:Tibshirani:1990} allows to model the dependence of the response variable in a flexible way using smooth functions of the predictors by defining the linear predictor as
$$
\eta_t = \beta_0 + \sum_{j=1}^p f_j(x_{tj}),
$$
where $f_j(x_{tj}) = \sum_{k=1}^{K_j} \beta_{jk} B_{jk}(x_{tj})$ is the smoothing term for the $j$th predictor with $\{B_{jk}()\}_{k=1}^{K_j}$ a set of known basis functions associated to unknown parameters $\beta_{jk}$.

Several smoothers can be defined by adopting different basis functions, such as penalized regression splines, cubic regression splines, etc. For an overview of the several smoothing functions available using splines bases see \citet[Chapter 5]{Wood:2017}.
Among the various possibilities, thin plate regression splines \citep[TPRS;][]{Wood:2003} represents a convenient form because TPRS (i) do not require to specify the ``knots'', (ii) use a low rank approximation of the full basis expansion, and (iii) are isotropic smoothers, so they are unaffected by any rotation or reflection of the covariates. 

In our application the only feature included as smoothing term in the linear predictor is time, so $x_{1}$ is an integer counting the days since the first day of the analysis. To some extent, the coding of such feature has no practical consequence, and other equivalent forms could have been used as well. 
In addition, to account for the reduced tracing activity during weekends (Saturday and Sunday) and holidays, a dummy variable $x_{2}$ is included taking value 1 for data referring to weekends or holidays, and 0 otherwise. 
The rationale behind the inclusion of such term is that the number of swabs processed is noticeably limited during weekends and holidays, so a significant increase in the test positivity rate is often observed due to the limited testing capacity and the higher probability of testing only symptomatic cases.

Thus, in our case the linear predictor of GAM simplifies to
$$
\eta_t = \beta_0 + \sum_{k=1}^K \beta_{1k} B_{1k}(x_{t1}) + \beta_2 x_{t2},
$$
where $\{B_{1k}\}_{k=1}^K$ represents the basis of thin plate regression splines. Note, however, that other smooth functions would have given nearly equivalent results.

\subsection{Estimation}
\label{estimation}

Estimation of the GAM model introduced in previous section can be pursued by REstricted Maximum Likelihood (REML), which amounts to maximize the penalized log-likelihood
\begin{equation}
\ell_P(\betab) = \ell(\betab) - \frac{1}{2} \lambda\betab\T\S\betab,
\label{eq:penloglik}
\end{equation}
where $\ell(\betab) = \sum_{t=1}^n \ell(y_t | \betab)$ is the log-likelihood for the observed values $y_t$ of the response variable.
The last term in the right-hand side represents the smoothing penalty, with $\lambda$ a smoothing parameter and $\S$ a known penalty matrix.

As reported in \citet{Wood:2011} and \citet{Wood:Pya:Saefken:2016}, REML is equivalent to marginal likelihood estimation of $\betab$ when the model contains Gaussian random effects, and it also leads to more stable estimates of $\lambda$ with much reduced risk of under-smoothing compared to GCV.
Furthermore, as discussed in the next section, the REML estimates of regression coefficients have an asymptotically MAP Bayesian interpretation that is very useful for obtaining simulated credible intervals for predictions. 
For a recent review on inference and computation in GAMs see \citet{Wood:2020}.

The selection of the smoothing parameter can be obtained, among many other proposals, by minimizing the conditional Akaike's information criterion (AIC). This version of AIC for GAMs uses the log-likelihood evaluated at the penalized MLE, and with the effective degrees of freedom computed as discussed in \citet{Wood:Pya:Saefken:2016}.

However, because the number of administered swabs is not constant over time, we must take into account this fact when modelling the test positive rate.
There are several reasons for this empirical evidence. First of all, during the weekends (particularly on Sundays) and holidays the number of swabs drops drastically. Furthermore, during periods of strong expansion of the pandemic, the monitoring system is unable to carry out effective surveillance and only symptomatic patients are likely to be tested.
Accounting for the different number of swabs in the model for the positive rate can be achieved by adopting a weighted penalized log-likelihood criterion. This amounts to replace the log-likelihood $\ell(\betab)$ in \eqref{eq:penloglik} with the weighted version
$$
\ell_W(\betab) = \sum_{t=1}^n w_t \ell(y_t | \betab),
$$
where $w_t$ are prior weights specifying the contribution of each data point to the log-likelihood.
In particular, indicating with $\bar{T}$ the average number of administered swabs over the period, weights can be defined as $w_t = T_t/\bar{T}$ so that positive rates $y_t$ computed from number of swabs larger than the average have proportionally larger weights, and vice versa for those rates based on number of swabs smaller than the average.
Furthermore, with the adopted definition for the weights the contribution of each datum is specified without changing the overall magnitude of the log-likelihood.

Once the model is fitted, the predicted TPR can be computed as
\begin{equation}
\hat{\mu}_t = \logistic\left( \hat{\beta}_0 + \sum_{k=1}^K \hat{\beta}_{1k} B_{1k}(x_{t1}) + \hat{\beta}_2 x_{t2} \right).
\label{eq:predmu}
\end{equation}
On certain occasions, for instance when computing the COVINDEX discussed in Section~\ref{covindex-as-a-monitoring-and-decision-making-tool}, we may want to compute predictions for the TPR with the weekends/holidays effect ruled out. This is easily accomplished by setting $x_{t2}=0$ for all $t$.

\subsection{Uncertainty and inference}
\label{uncertainty-and-inference}

The penalized likelihood approach described above has also a Bayesian interpretation by assuming an improper multivariate normal prior on $\betab$. In this case, the REML estimates of $\betab$ coefficients are asymptotically the maximum a posteriori (MAP) of the Bayesian posterior distribution, with the latter given by
\begin{equation}
\betab | (\y, \lambda) \sim N(\hat{\betab}, (\hat{\Info} + \lambda\S)^{-1}),
\label{eq:coef_posterior}
\end{equation}
where $\hat{\Info}$ is the observed information matrix (Hessian of the negative log-likelihood) at $\hat{\betab}$ \citep[Section 6.10]{Wood:2017}.
This result is useful for computing approximate credible intervals for any function of $\betab$ by simulating from the posterior \citep[Section 7.2]{Gelman:Hill:2007}. \citet[p. 294]{Wood:2017} reported good frequentist coverage properties for such Bayesian credible intervals, with empirical coverage close to the nominal level when averaged across the domain of the function.

In practice, coefficients $\betab^*$ are simulated from \eqref{eq:coef_posterior}, and then plugged in equation \eqref{eq:predmu} to get the simulated means $\mu_t^*$. The process is replicated a large number of times, say $10\,000$ or more, and the percentiles of the simulated distributions at different values of $x_t$ can be used to compute the limits of approximate credible intervals for the mean. 
To compute approximate credible intervals for the single prediction we simulate response values as $y^*_t \sim \Beta(\mu^*_t, \hat{\phi})$, where $\mu_t^*$ is the simulated mean as described above, $\hat{\phi}$ is the model estimate of the precision parameter, and then we compute the percentiles of the simulated distribution of predicted values for the response. The empirical coverage of the prediction intervals will be assessed in Section~\ref{gambetareg-estimate}.

\section{COVINDEX as a Monitoring and Decision-Making Tool}
\label{covindex-as-a-monitoring-and-decision-making-tool}

The COVINDEX proposed in this paper is an attempt to compute a synthetic index summarizing the evolution of the COVID-19 pandemic, which can be useful to policy makers and public health officials for monitoring local and national outbreaks.
In our proposal this is simply computed as
\begin{equation}
\covindex_t = \frac{\hat{\mu}_t }{\hat{\mu}_{t-7}},
\label{eq:predcovindex}
\end{equation}
the ratio of the predicted positive rate at time $t$ to the prediction 7 days earlier. The value of 7 is chosen because it is approximately the expected incubation time for COVID-19 \citep{Nazar:Elfadil:2021}, and because it corresponds to the observed weekly fluctuation in testing.
A COVINDEX value larger than 1.0 means that the pandemic is growing, while a value smaller than 1.0 indicates that new infections are slowing down.

The COVINDEX estimate is clearly affected by uncertainty and to account for it the approach outlined in Section \ref{uncertainty-and-inference} for TPR can be used here as well.
In particular, for each simulated series of values $\mu_t^*$, simulated COVINDEX series can be obtained by applying equation \eqref{eq:predcovindex} to get the simulated values
$\covindex^*_t = \mu^*_t / \mu^*_{t-7}$. Approximate credible intervals can then be computed from the percentiles of the simulated distribution.

\begin{figure}[htb]
\centering
\includegraphics[width=0.6\linewidth]{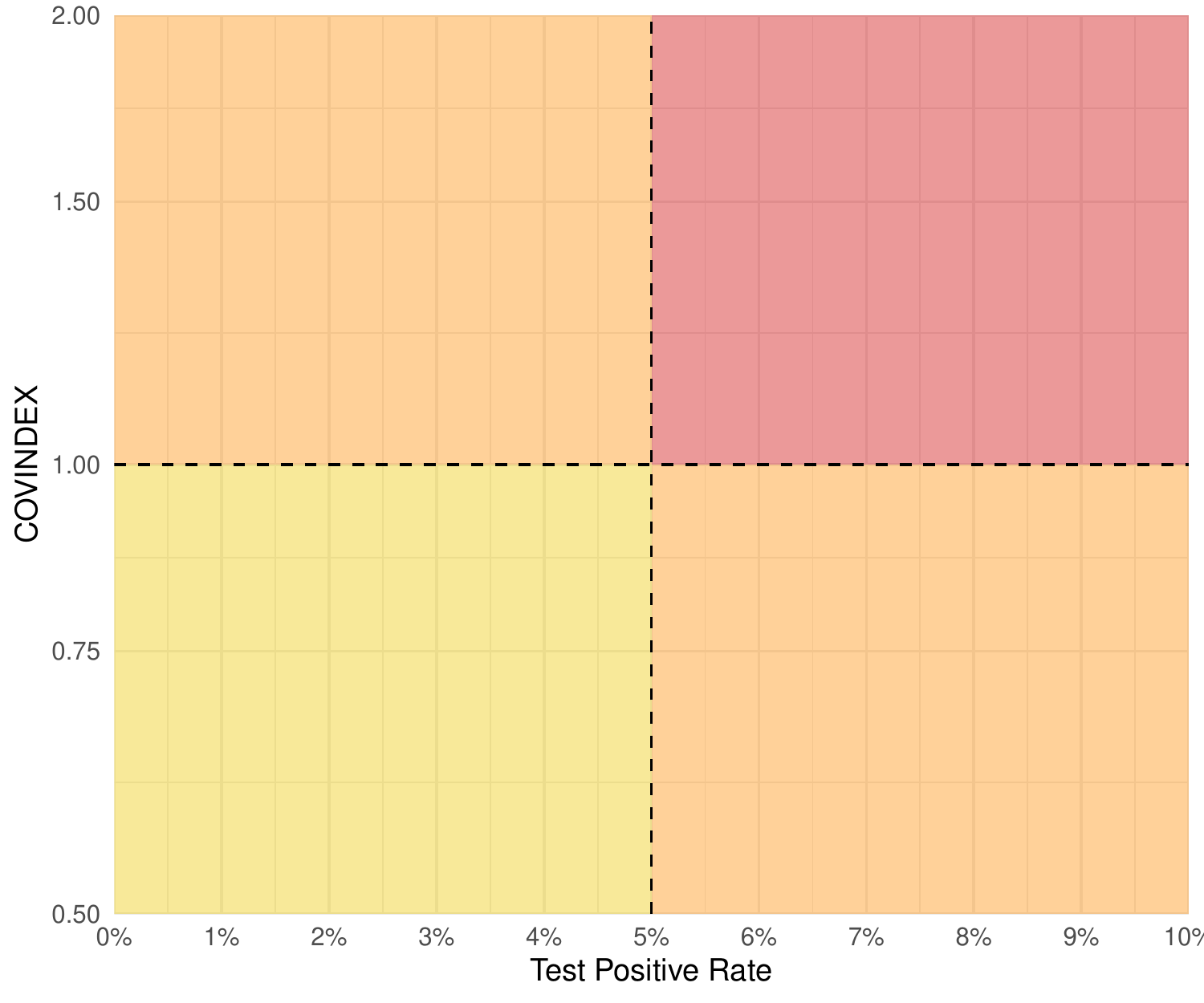} 
\caption{TPR-COVINDEX risk quadrant chart.}
\label{fig:quadrants}
\end{figure}

We argue that decisions made by policy makers should be based both on the COVINDEX, which provides an outlook on the likely behaviour of the pandemic in the near future, and on the level of the estimated TPR, which represents its current status.
Following this idea, a TPR-COVINDEX risk quadrant chart can be drawn (see Figure \ref{fig:quadrants}). This chart illustrates four potential scenarios which represent a useful tool for a decision maker.
The quadrants are defined by the dashed lines drawn at selected threshold values.
For COVINDEX the natural reference value is 1.0, with values below it indicating a shrinking outbreak, and values higher than 1.0 indicating epidemic situations that are increasingly worrying and out of control.
Note that, since the index is a ratio, the $y$-axis is expressed in logarithmic scale.
For the positive rate, the threshold value can be set according to the World Health Organization, which published a set of criteria to inform whether the epidemic is under control. In particular, one criterion states that ``{[}\ldots{]} less than 5\% of samples positive for COVID-19, at least for the last 2 weeks, assuming that surveillance for suspected cases is comprehensive'' \citep{WHO:2019}.

According to the above mentioned threshold values, the upper-right quadrant represents the worst-case scenario, with high values of both TPR and COVINDEX. On the contrary, the best-case scenario is the lower-left quadrant which has both low TPR and COVINDEX less than 1.0 indicating a decreasing circulation of the virus. The remaining quadrants are intermediate cases. Typical situations will move in a clockwise direction, moving from the worst-case, represented by the red quadrant on top-right, to the orange quadrant at bottom-right, and eventually reaching the yellow quadrant indicating an outbreak under control.
However, in some cases the pandemic could regain strength by getting COVINDEX values greater than 1.0, thus moving towards the top-left orange quadrant or directly towards the worst-case situation described by the red quadrant.
A description of the Italian situation since March 2020 is discussed in Section \ref{application-to-italian-covid-19-pandemic}.

\section{Application to Italian COVID-19 Pandemic}
\label{application-to-italian-covid-19-pandemic}

\label{sec:ita}

\subsection{Data}
\label{data}

The Italian Department of Protezione Civile provides daily information on the COVID-19 pandemic, both at the national and the regional level, in a public GitHub repository \citep{DPC:COVID19}.
Among the data contained in this repository, the cumulative number of naso-pharyngeal or molecular swabs and the corresponding positive tests are provided.
Starting with January 15th, 2021, antigen tests are also officially recorded, while previously only some regions included them in the recorded statistics since autumn 2020.
The reliability of such information is at best questionable and not available uniformly for the year 2020.
For these reasons, in our analyses we considered the information from daily molecular swabs (not persons tested) to compute the test positive rate (TPR), a commonly used screening and diagnostic tool for COVID-19 \citep{WHO:2020}. 

\begin{figure}[htb]
\centering 
\includegraphics[width=\linewidth]{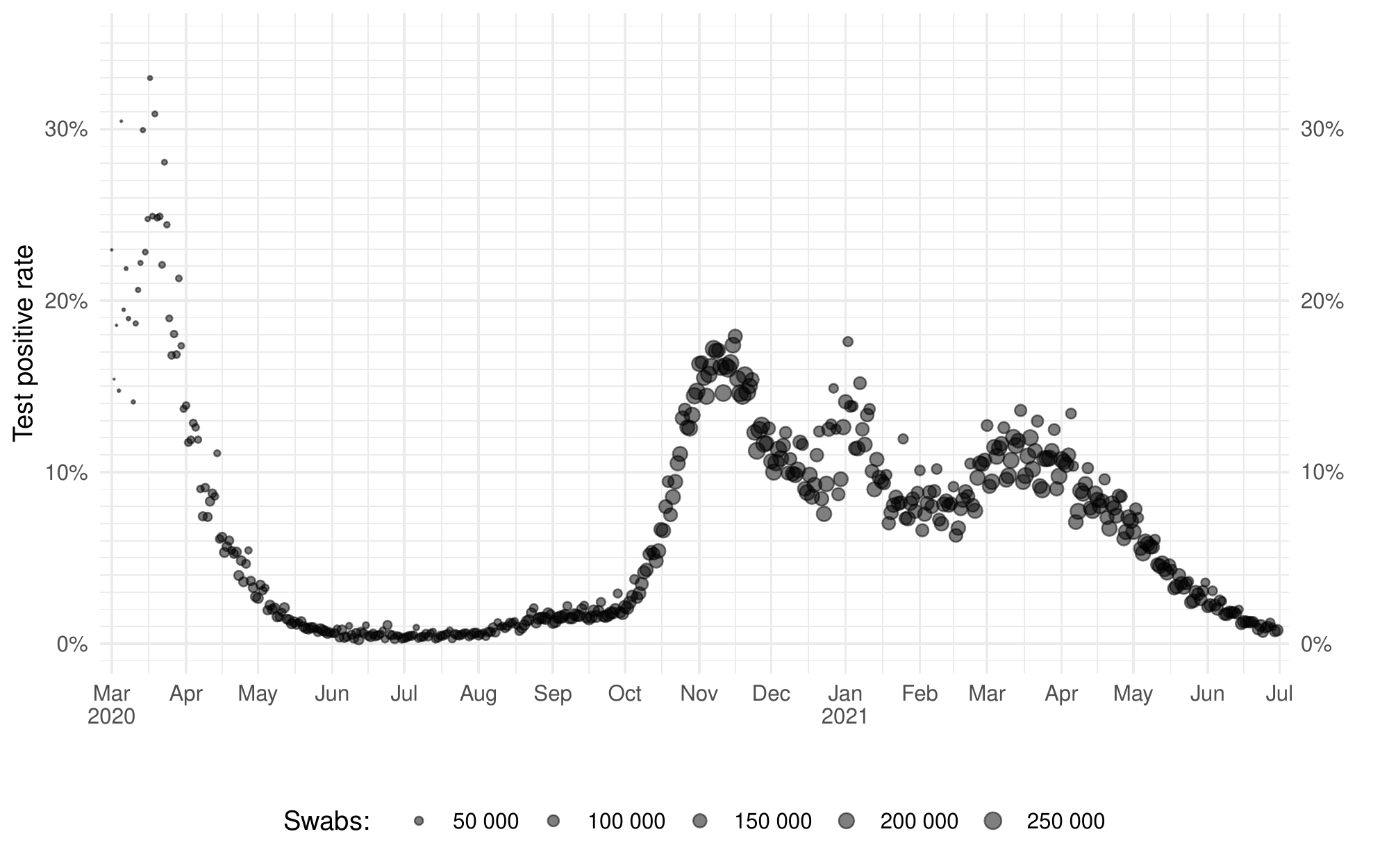} 
\caption{Plot of test positive rate from beginning of COVID-19 pandemic in Italy to the end of observational period with size of points proportional to the number of molecular swabs administered.}
\label{fig:plot-data}
\end{figure}

The plot on Figure \ref{fig:plot-data} shows the observed TPR over time with points proportional to the administered swabs.

\subsection{GAM beta regression model estimate}
\label{gambetareg-estimate}

\begin{figure}[htb]
\centering 
\includegraphics[width=0.8\linewidth]{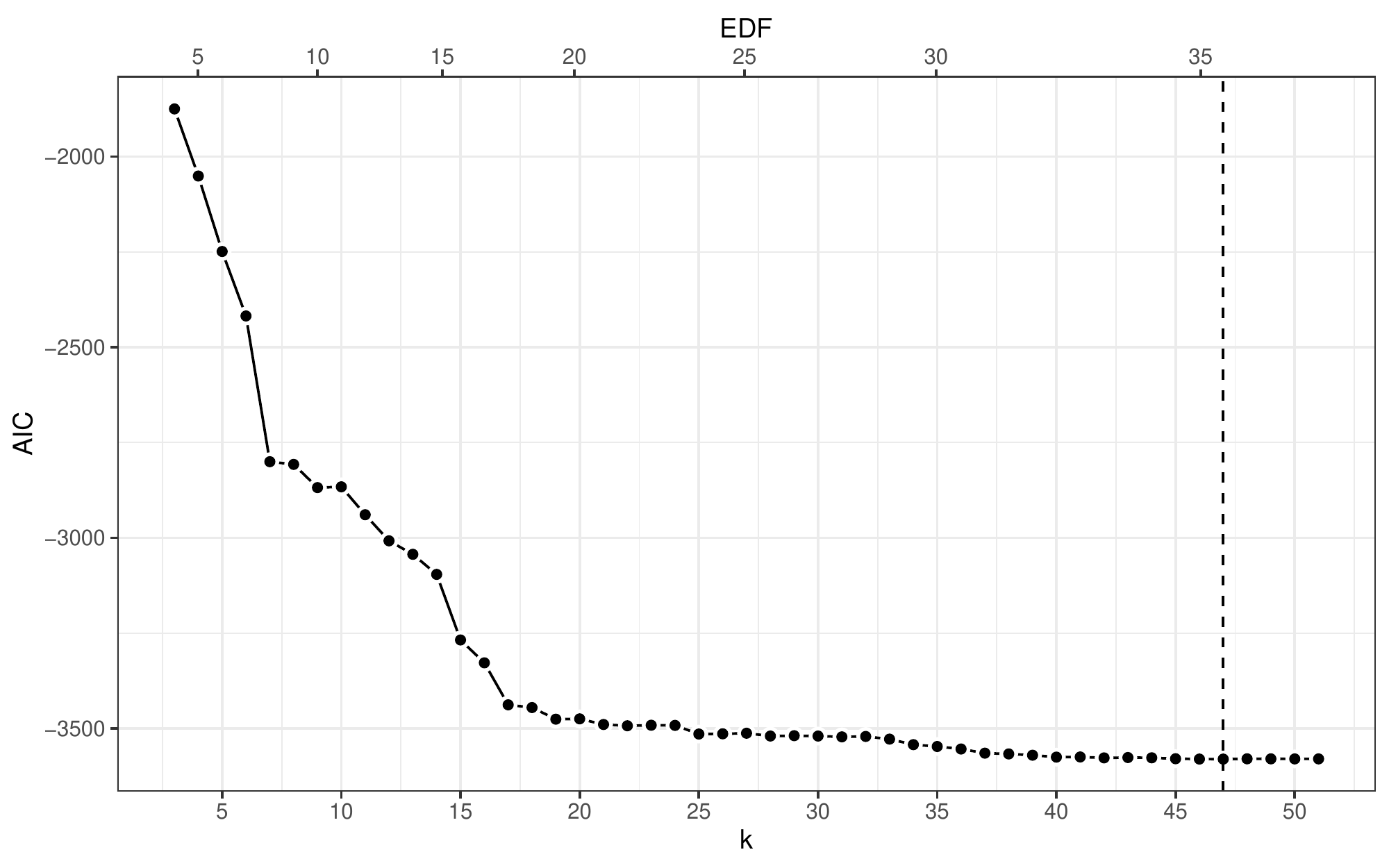}
\caption{Trace plot of AIC as a function of $k$, the number of basis functions used by the thin plate regression smoother, with the corresponding effective degrees of freedom (EDF). The EDF expresses the complexity of the smoother, with larger values indicating more wiggly smoothers, and it cannot be larger than $k$. The vertical dashed line is drawn at the minimum AIC used for the selection of the final model.}
\label{fig:aic}
\end{figure}

\begin{figure}[htb]
\centering 
\includegraphics[width=\linewidth]{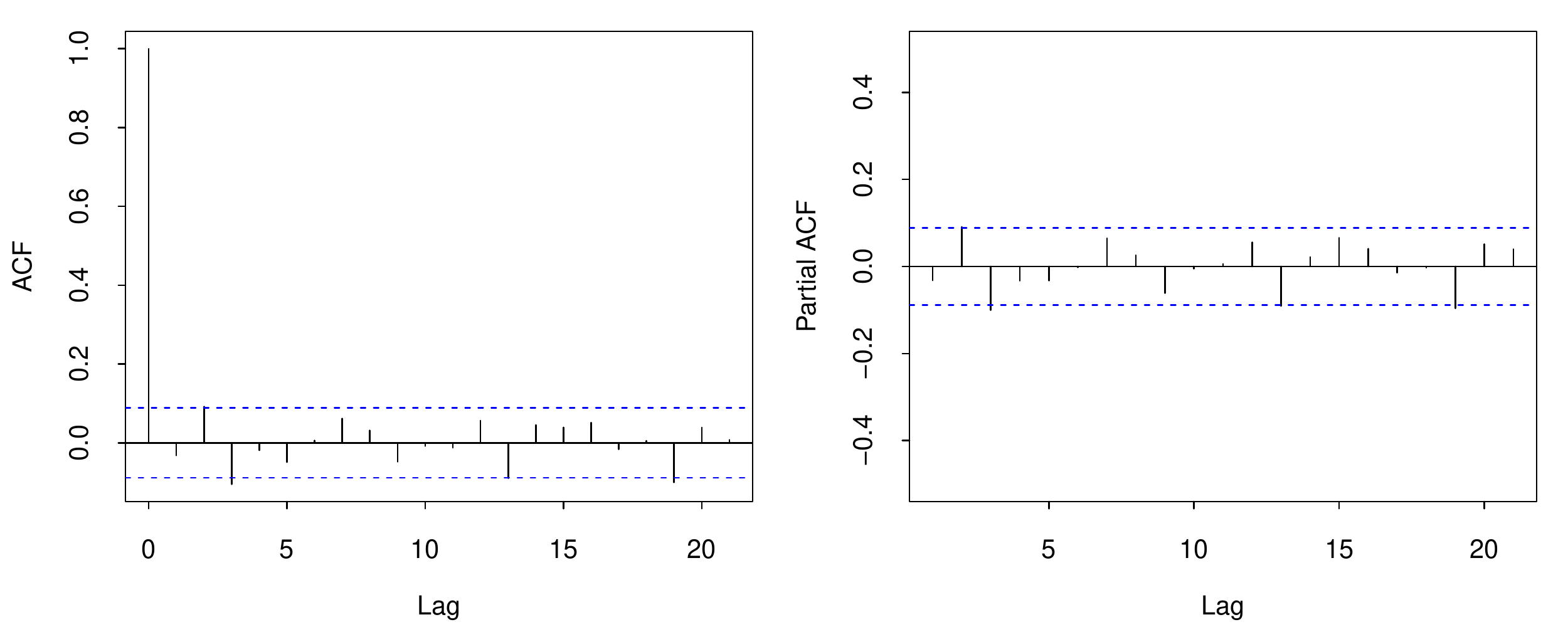} 
\caption{Autocorrelation and partial autocorrelation functions for the deviance residuals of the estimated GAM beta regression in Table~\ref{tab:gam_summary}}
\label{fig:acf_pacf}
\end{figure}

Table \ref{tab:gam_summary} reports the summary output of the estimated GAM beta regression model for the test positive rate in Italy from March 1st 2020 to June 30th 2021. The parametric terms include the intercept and a dummy variable for the days following the weekends (Saturday and Sunday) and holidays.
The smooth term captures the evolution of underlying trend in the observed test positive rate. The amount of smoothing applied to the time predictor is selected by minimizing the AIC, as shown in Figure~\ref{fig:aic}. The graphs of the autocorrelation and partial autocorrelation functions for the deviance residuals in Figure~\ref{fig:acf_pacf} show no significant remaining correlation at different lags.

\begin{table}[ht]
\centering
\caption{GAM beta regression model summary.}
\label{tab:gam_summary}
\begin{tabular}{lrrrr}
\toprule
\multicolumn{2}{l}{Num. of obs. =  487 } & 
     \multicolumn{3}{r}{Dispersion par. =  1467.3 }\\
\multicolumn{2}{l}{Log-likelihood =  1828.7 } & 
     \multicolumn{3}{r}{Deviance expl. =  0.9891 }\\
\multicolumn{2}{l}{REML =  1726.3 } &
     \multicolumn{3}{r}{AIC =  -3583.1}\\
\midrule\multicolumn{5}{l}{Parametric coefficients:} \\
     & Estimate & Std. error & z-value & $p$-value \\ 
(Intercept) & -3.1904 & 0.01177 & -271.07 & < 0.001 \\
Weekend     &  0.1681 & 0.01023 & 16.44 & < 0.001 \\
\midrule\multicolumn{5}{l}{Smooth terms:} \\
     & edf & Ref. df & ChiSq-value & $p$-value \\ 
$s(t)$ & 35.11 & 39.67 & 16820 & < 0.001 \\
\bottomrule
\end{tabular}
\end{table}

\begin{figure}[htb]
\centering 
\includegraphics[width=\linewidth]{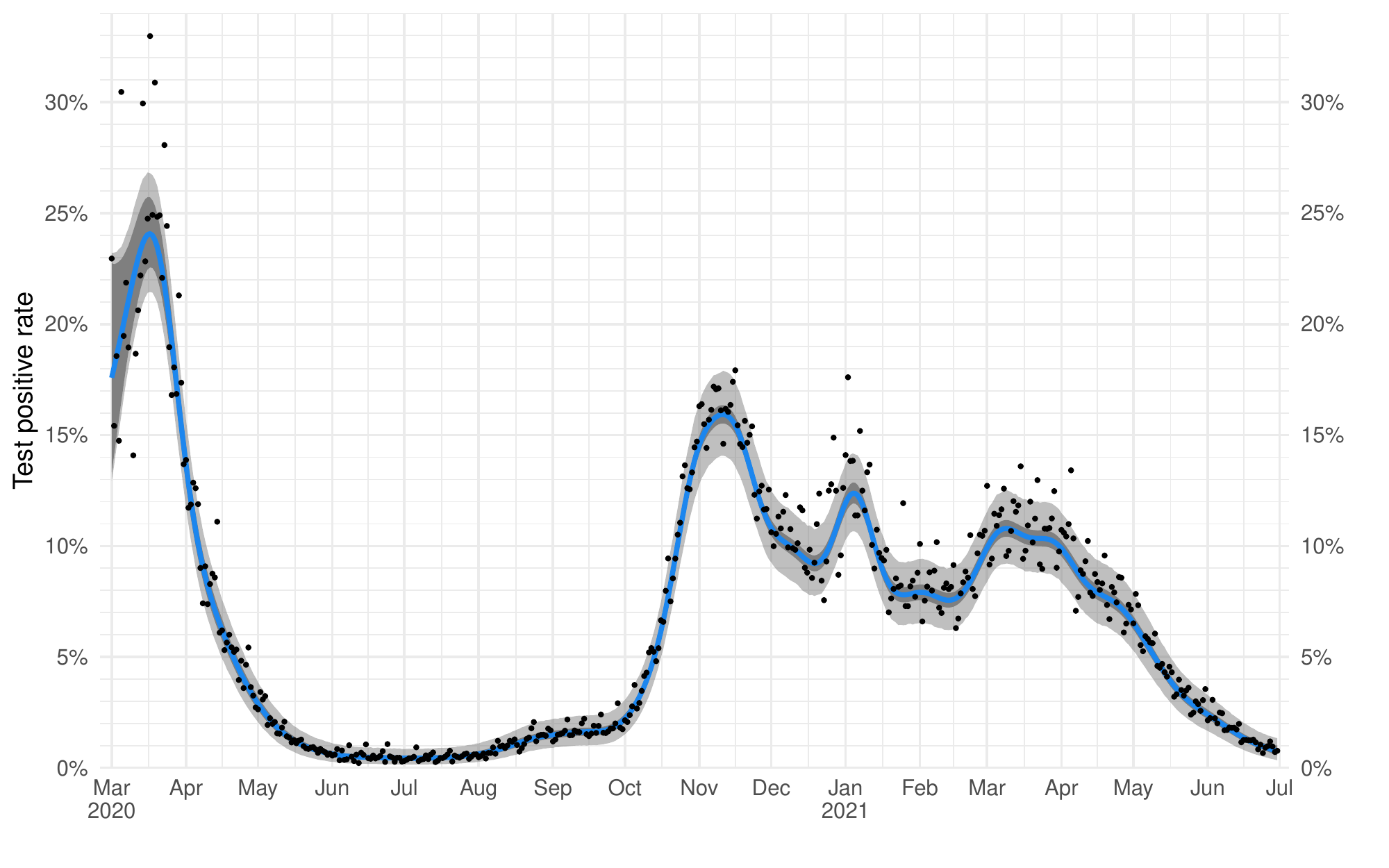} 
\caption{GAM estimate of test positive rate (blue line) in Italy during 2020 and first half of 2021, with 95\% simulated credible intervals for the mean (dark grey area) and for the single value (light grey area).}
\label{fig:TPR-Italy}
\end{figure}

Figure \ref{fig:TPR-Italy} reports the estimated curve for the test positive rate with 95\% credible intervals for the mean and the single value obtained by simulating from the posterior distribution as described in Section \ref{uncertainty-and-inference}.
The highest positive rates are achieved in March 2020 during the first wave of pandemic, and on November 2020, corresponding to the second wave. 
A resurgence of spread during the end of 2020 is followed by a quick decrease in earlier 2021. 
Subsequently, the situation remained stable for about a month, but in the second part of February another sharpe increase occurred due to the appearance of COVID-19 variants in the Italian territory (in particular the Alpha or english variant). 
Starting from the beginning of April, a marked decline in the TPR can be observed, likely favored by the increased full vaccination coverage of the Italian population. 

\subsection{Empirical coverage of credible intervals for predictions}

\begin{figure}[htb]
\centering 
\includegraphics[width=\linewidth]{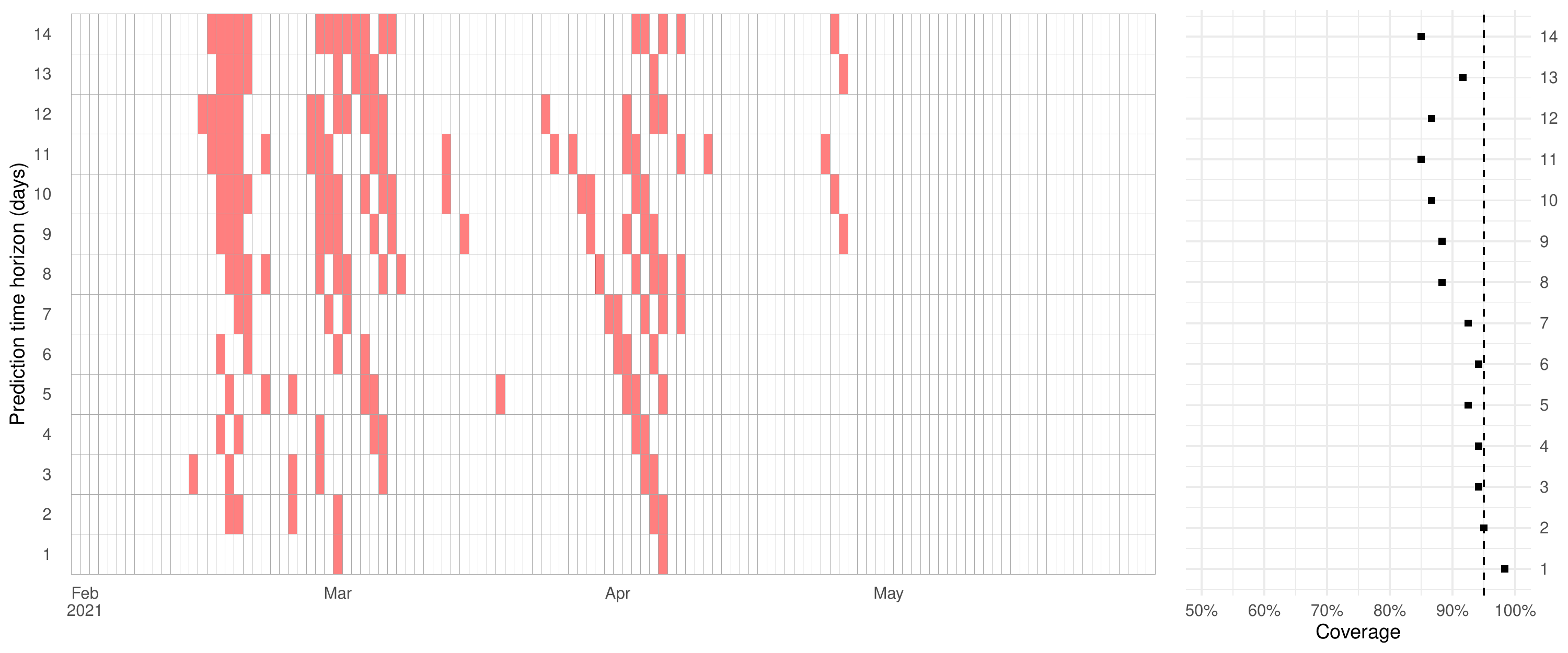} 
\caption{Prediction intervals coverage for the GAM beta regression model of TPR at 95\% nominal level. The graph on the left shows the inclusion (blank cells) or exclusion (red cells) of the observed value of TPR for the prediction intervals in subsequent dates corresponding to the time horizon from 1 up to 14 days. The empirical coverage percentages are shown on the right graph, with the vertical dashed line representing the nominal level.}
\label{fig:TPR-coverage}
\end{figure}

To investigate the accuracy of the simulated prediction intervals, we considered all the dates from February 1st 2021 to May 31st 2021 and the time horizon for the predictions from 1 day to 14 days. 
For each date we fitted the GAM beta regression model using all the data available up to that day, and then we used the estimated model to compute the simulated credible intervals for the following 14 days. 
This process was replicated for the all the days in the specified range and the empirical coverage calculated. 
Figure~\ref{fig:TPR-coverage} reports on the left panel a graph showing the inclusion or exclusion of the observed values of TPR in the simulated prediction intervals, while on the right a plot of the empirical coverage for the time horizon from 1 up to 14 days. 
Overall the coverage is close to the nominal level, with all the values above 90\% for the forecasts of the first week, and between 85\% and 90\% for the second week.
It is interest to note that most of the coverage errors occur in periods of abrupt changes, for instance at the sharp rise of TPR in the last week of February or at the beginning of TPR decline in mid-March. As expected, the empirical coverage decreases as the prediction horizon increases.

\subsection{COVINDEX estimate}
\label{covindex-estimate}

\begin{figure}[htb]
\centering 
\includegraphics[width=\linewidth]{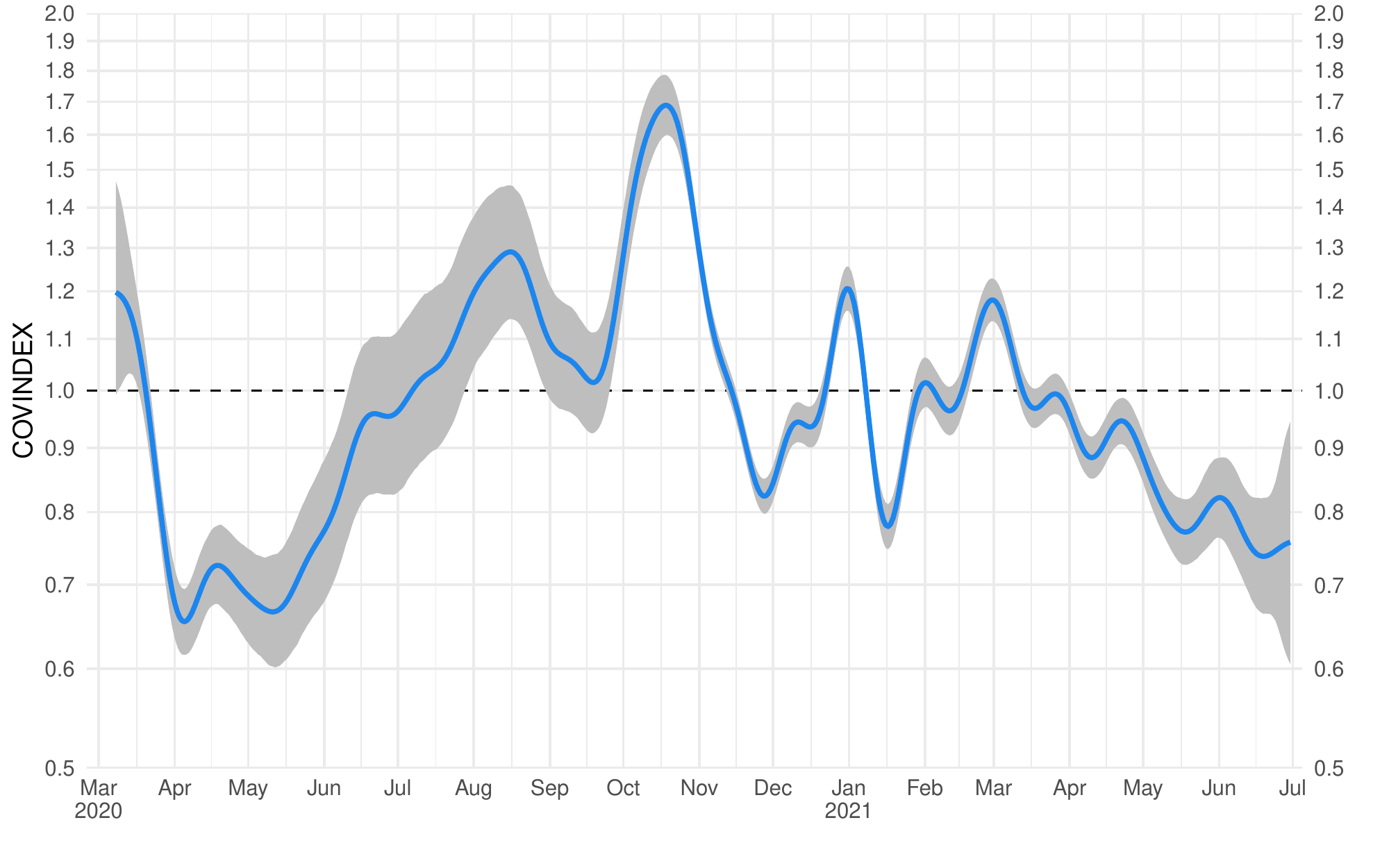} 
\caption{Evolution of COVINDEX (on logarithmic scale) for Italy during 2020 and first half of 2021, with approximate 95\% simulated credible intervals.}
\label{fig:COVINDEX-Italy}
\end{figure}

Based on the estimated model and uncertainty for the test positive rate, the COVINDEX is computed following equation \eqref{eq:predcovindex}. Figure \ref{fig:COVINDEX-Italy} shows the estimated COVINDEX with 95\% credible intervals. Notice that the $y$-axis is expressed on logarithmic scale, the natural scale to visualize ratios \citep[][, Sec. 3.2]{Wilke:2019}.
The index fluctuates widely throughout the year 2020, following the periods of expansion and contraction of the spread of the pandemic. After the first wave in spring 2020 we observe a quick decreasing trend, followed by a slowly increase during the summer, corresponding to a relaxation of the containment measures, with values significantly larger than 1.0 during August. This represents the first signal of a resurgence of the pandemic. Sharpe and large increases are also observed during October in conjunction with the second wave that strongly affected Italy.
Two additional peaks are detected at the end of 2020 and on February 2021, corresponding to gradual relaxation of containment measures before Christmas and mid January, with the latter that occurred during the period of political instability associated with the change of government.

\begin{figure}[htb]
\centering 
\includegraphics[width=\linewidth]{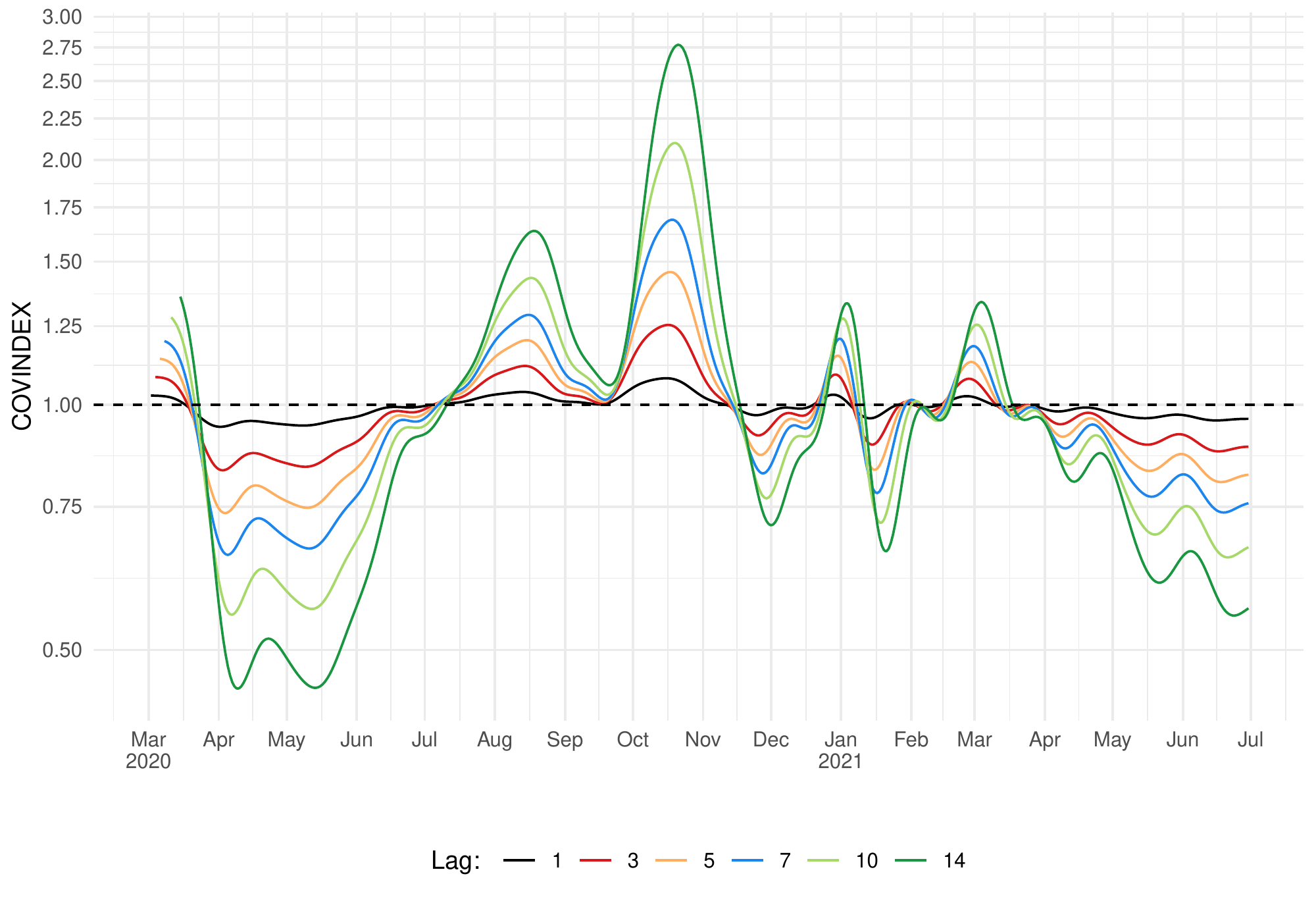} 
\caption{A comparison of COVINDEX (on logarithmic scale) computed at different lags for Italy  on 2020 and first half of 2021.}
\label{fig:COVINDEX-lag}
\end{figure}

From the adopted definition of equation~\eqref{eq:predcovindex}, COVINDEX is computed by taking the ratio of the estimated TPR with respect to a 7-days-before estimate. In Section~\ref{covindex-as-a-monitoring-and-decision-making-tool} we provide the rationale for this choice. However, it may be interesting to investigate how the index changes assuming different lags. Figure~\ref{fig:COVINDEX-lag} shows the COVINDEX estimates obtained when different lag values are used. 
The general behaviour of the curves is similar across different lags, but the amplitude of the oscillation increases as the lag increases. 
This appears reasonable since, essentially, COVINDEX compares the estimated TPR at time $t$ with the value at time $t-\text{lag}$. Thus, for smaller values of the lag the index fluctuates less and is more stable than at higher lag values. 
However, lag values that are too small cannot highlight the dynamics of TPR because too close values tend to be quite similar. In this sense, the selected 7-day lag appears to be a sensible choice. 
 
\subsection{TPR-COVINDEX risk quadrant chart analysis}
\label{tpr-covindex-risk-quadrant-chart-analysis}

\begin{figure}[htb]
\centering
\includegraphics[width=0.7\linewidth]{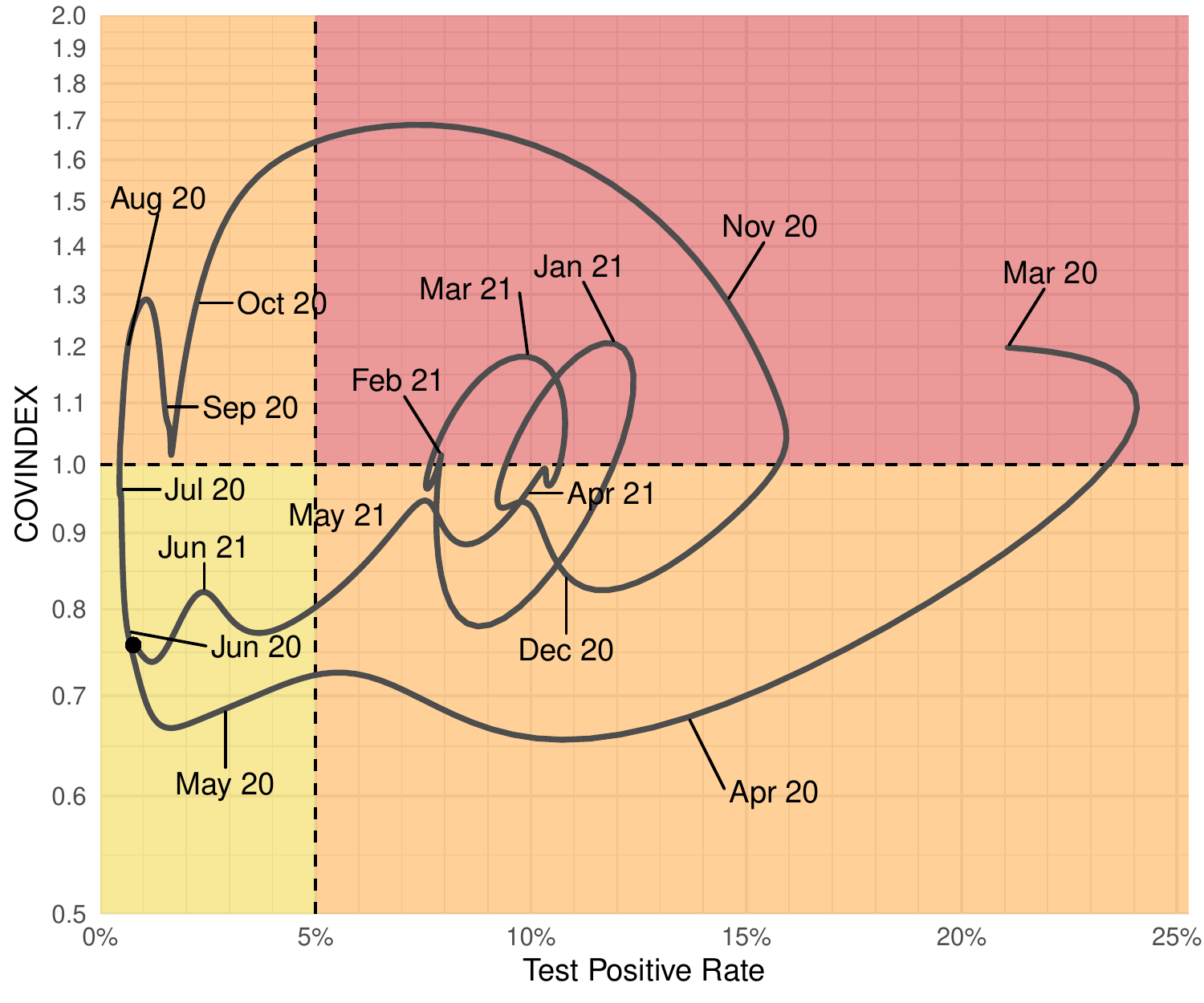} 
\caption{TPR-COVINDEX risk chart as connected scatterplot for Italy. The first day of each month is highlighted to provide a temporal reference.}
\label{fig:quadrants-Italy}
\end{figure}

Figure \ref{fig:quadrants-Italy} shows the TPR-COVINDEX risk quadrant chart for Italy, with points connected following the temporal path, a graph also known as connected scatterplot \citep{Haroz:etal:2015}. The curve concisely represents the evolution of both indices during the pandemic.
Starting with the critical situation in March 2020, the situation improved in the following months, moving from the red quadrant to the orange bottom-right quadrant and then the yellow quadrant during summer 2020. By the end of summer 2020 we observe a worsening of the situation that lead to the red quadrant in November. In the following months there has been a constant oscillation between the red and right-orange quadrants, indicating a serious pandemic situation.

\begin{figure}[htb]
\centering
\includegraphics[width=1\linewidth]{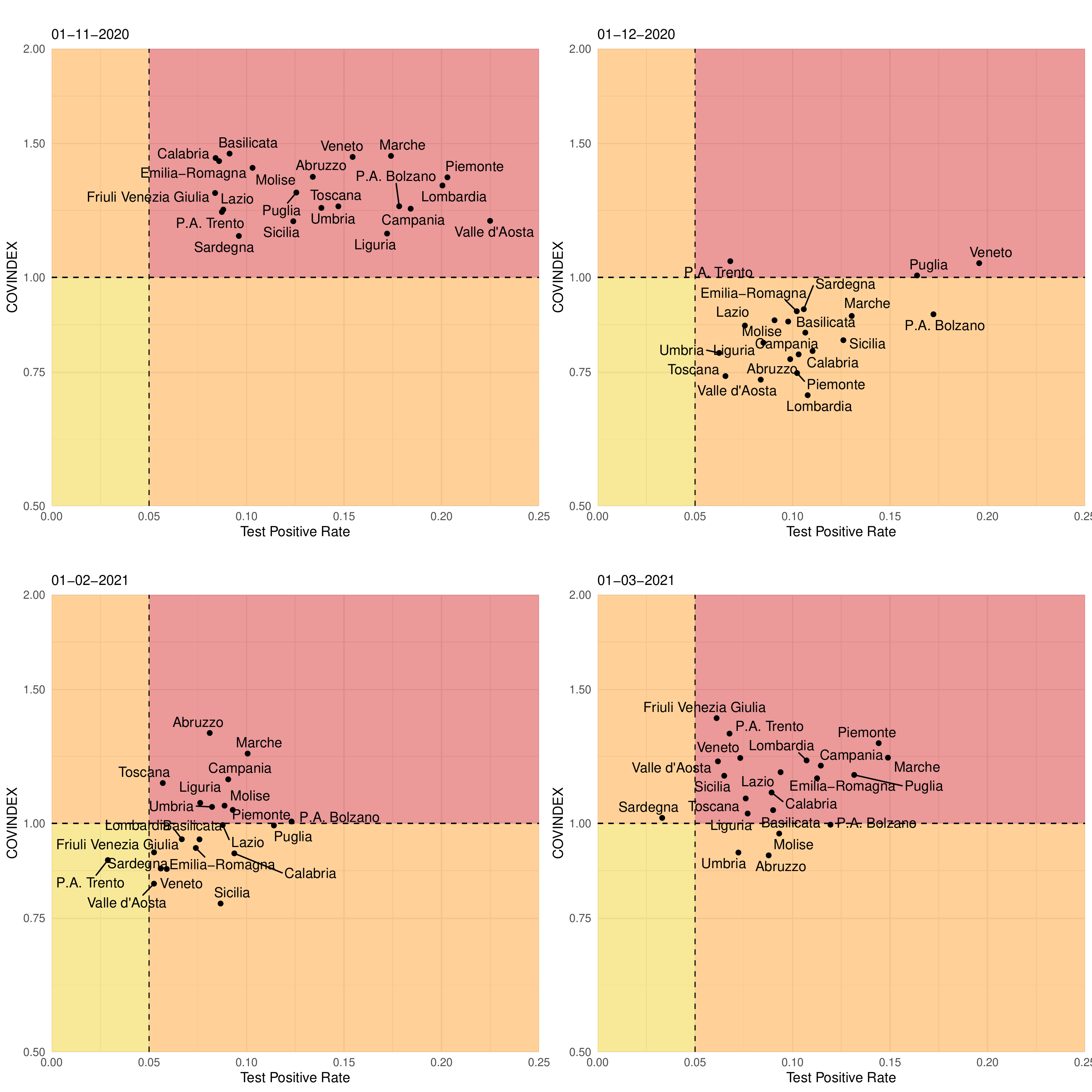} 
\caption{TPR-COVINDEX plot for Italian regions at different time points.}\label{fig:quadrants-regions}
\end{figure}

A scatterplot of TPR vs COVINDEX is also useful for surveillance of the pandemic in different Italian regions. Figure \ref{fig:quadrants-regions} summarizes the status of the pandemic for the Italian regions at selected time points.
A high-risk situation is observed at the beginning of November 2020, where all regions belong to the red quadrant. The following month saw an improvement with most regions moving towards the bottom-right orange quadrant.
A more complex and varied situation is observed between February and March 2021, with some regions moving from the red to the orange quadrant, and vice versa for other regions.

\clearpage

\section{A comparison of COVINDEX with the effective reproduction number}
\label{a-comparison-of-covindex-with-the-effective-reproduction-number}

The main index used in Italy for pandemic surveillance is $R_t$, the effective reproduction number. The procedure employed for estimating $R_t$ is described by \citet{Guzzetta:Merler:2020} and it is based on the Bayesian methodology of \citet{Cori:etal:2013}.
Details can be found at \url{https://www.epicentro.iss.it/coronavirus/sars-cov-2-sorveglianza-dati}. An archive containing both the data and the R script used by the Italian National Institute of Health (ISS) for computing $R_t$ is available at \url{https://www.epicentro.iss.it/coronavirus/open-data/calcolo_rt_italia.zip}.

In this Section we provide a comparison of the proposed COVINDEX with the values of $R_t$ estimated following the procedure outlined above for Italy from March 2020 to June 2021.
Furthermore, since the effective reproduction number does not provide a timely snapshot of the evolution of the pandemic, we also provide two examples showing the failure of $R_t$ to highlight the likely evolution of the pandemic and we compare its behaviour with the proposed COVINDEX.

\begin{figure}[htb]
\centering
\includegraphics[width=\linewidth]{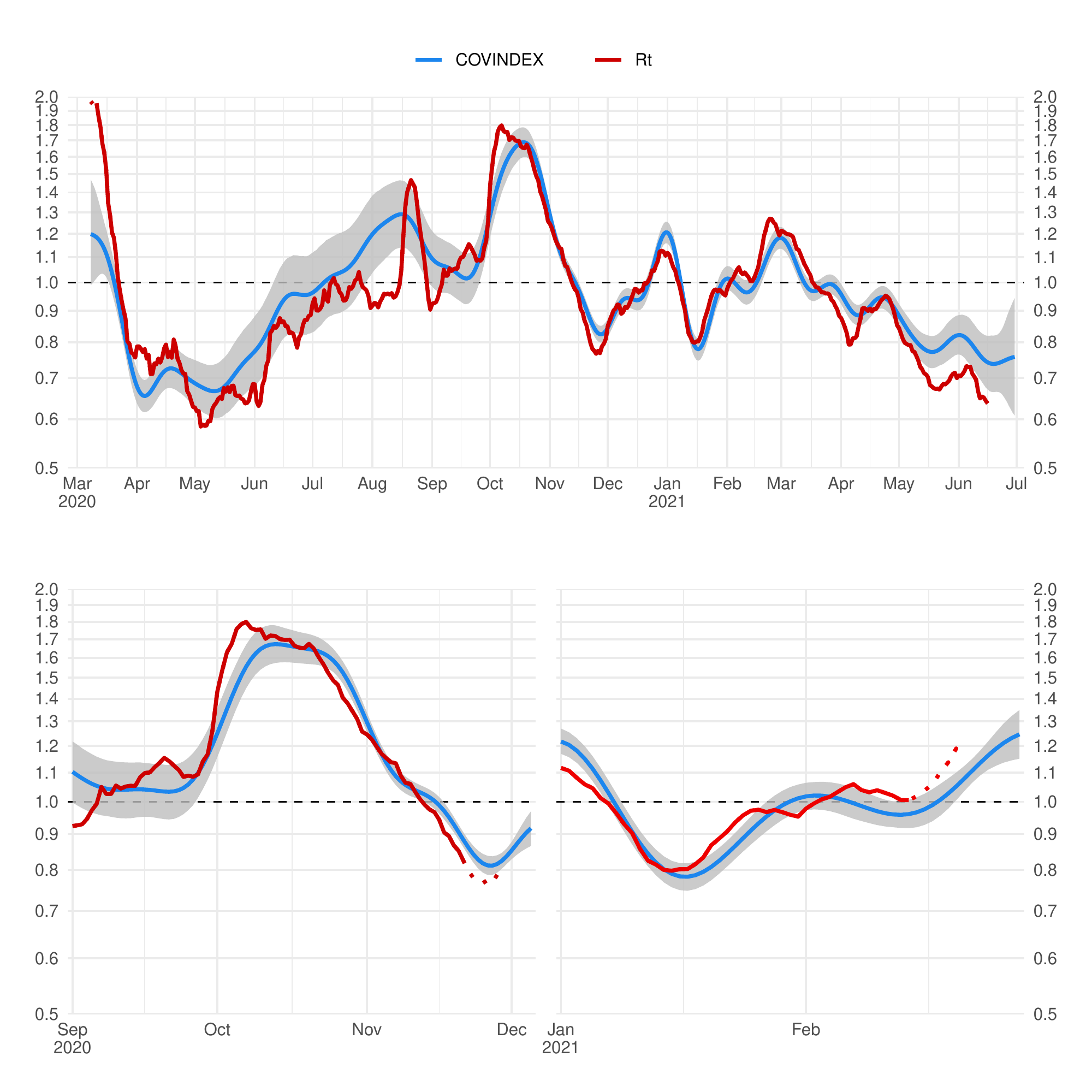}
\caption{Comparison of COVINDEX and $R_t$ for Italy. Bottom panels show the comparison at early December 2020 (left) and at the end of February 2021 (right).}
\label{fig:covindex-vs-Rt}
\end{figure}

The top graph reported in Figure \ref{fig:covindex-vs-Rt} shows the estimated curves for COVINDEX and $R_t$.
Overall, a similar trend can be observed for the two curves, particularly since October 2020. $R_t$ appears to be more wiggly than COVINDEX, especially during the summer 2020. Likely, this is related to the large uncertainty in that period due to the relative small number of positive cases (around few hundreds) observed in that period.
One of main drawbacks of using $R_t$ for real-time monitoring is shown in the final part of the graph. In fact, if at the end of the June the COVINDEX curve seems to suggest a resumption of the pandemic, the $R_t$ index continues to show a decreasing trend
This behaviour can also be seen in other time periods, as discussed below.

As mentioned in Section \ref{covindex-estimate}, the second wave of COVID-19 epidemic hit Italy between the second half of October and the beginning of November 2020, followed by a rapid decrease during the remainder of the month.
However, from the beginning of December 2020 it was evident that this decline had stopped and that the situation was starting to get worse.
This is clearly indicated by the upward slope of the COVINDEX computed on December 5th and shown in the bottom-left graph of Figure \ref{fig:covindex-vs-Rt}.
On the contrary, the $R_t$ index calculated on the same day, with estimates considered valid up to 14 days before, produces a curve which erroneously suggests a decline in the spread of the pandemic.
However, if the $R_t$ curve is estimated a week later, we begin to see an increase in the spread of the pandemic (see the dotted red curve in bottom-left graph of Figure \ref{fig:covindex-vs-Rt}). The main problem is that such alert is reported too late.

A similar situation is also faced at the beginning of March 2021. After a period of almost constant positive rate during February 2021, with both COVINDEX and $R_t$ oscillating around 1.0, by the end of the month there was a clear increase of the test positive rate. This was immediately signalled by COVINDEX computed on February 28th 2021 (see bottom-right graph in Figure \ref{fig:covindex-vs-Rt}), but $R_t$ computed on the same day was still signalling a steady state and only after a week an increasing value of $R_t$ would have signalled the resurgence of the pandemic.

The comparison between COVINDEX and $R_t$ can also be conducted at the regional level. Here we present a comparison for two Italian regions, Lombardia and Umbria. These are two very different regions, both in size and geographical position, but also in terms of pandemic history. 
If Lombardia was the most affected region of Italy during the 1st wave of the COVID-19 pandemic, Umbria suffered only marginal effects in this phase. 
On the contrary, the so-called 3rd wave that occurred in winter/spring 2021 hit Umbria earlier than in the rest of Italian regions, including Lombardia.

\begin{figure}[htb]
\centering
\includegraphics[width=0.8\linewidth]{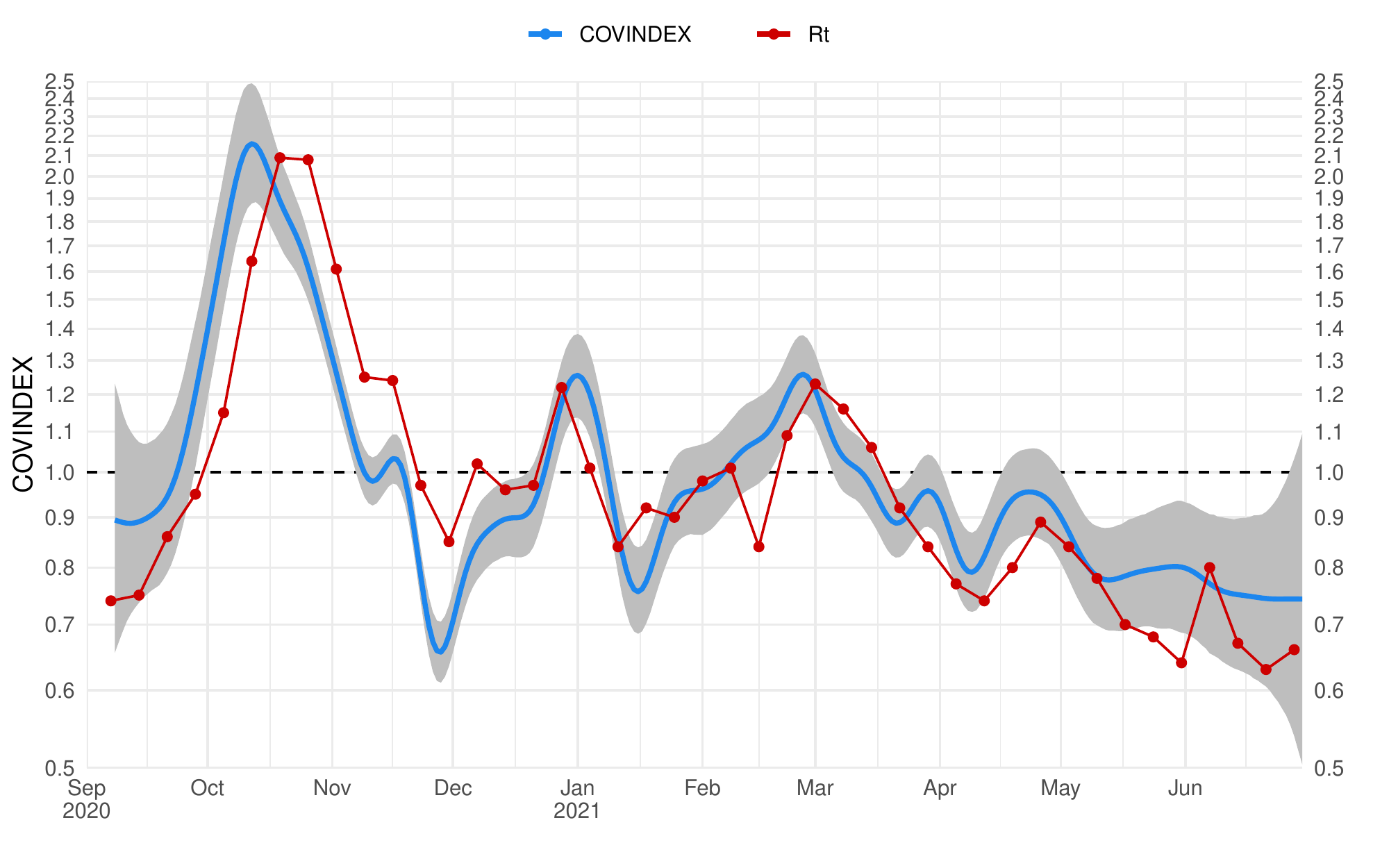}\\
\includegraphics[width=0.8\linewidth]{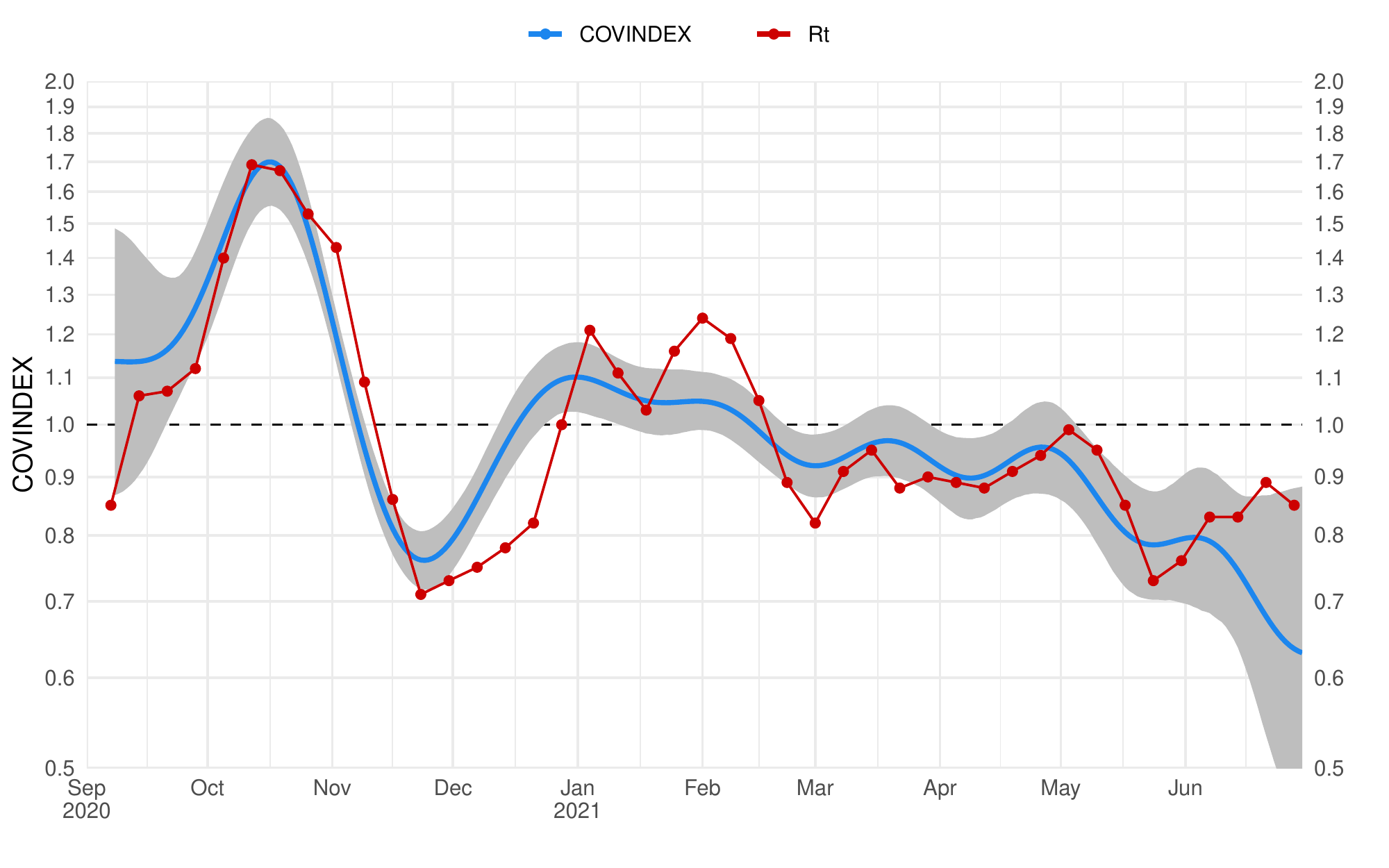}
\caption{Comparison of COVINDEX and $R_t$ for the Italian regions Lombardia (top) and Umbria (bottom), from September 2020 to June 2021.}
\label{fig:covindex-vs-Rt-regions}
\end{figure}

Likewise the national level, there is a substantial similarity between the trend of COVINDEX and $R_t$ for the two regions, with the former which appears to have a smoother behaviour (see Figure~\ref{fig:covindex-vs-Rt-regions}). Both indices correctly identified the peak of the pandemic in October 2020 and at the end of 2020. 
But if for Umbria the beginning of 2021 is marked by the arrival of the third wave caused by the circulation of SARS-CoV-2 variants, namely Alpha (or English) and Gamma (or Brazilian), in Lombardia the presence of these variants only occurred from mid-February. Subsequently, starting from spring 2021, a decline in the epidemic can be observed in both regions. 

However, there are also differences that are worth pointing out.
For Lombardia there are two values of $R_t$, in mid-February and early June, which appear suspicious as they are placed outside the apparent trend, underestimating in the first case and overestimating the trend in the second case. 
For Umbria, the $R_t$ seems to increase starting from the last week of May, while the COVINDEX still suggests a decreasing trend. This behaviour of $R_t$ is also suspect as the TPR of the region remains substantially stable or slightly decreasing in this period, with all the TPR values less than 1\% in the last 10 days of June.

\section{Final comments}
\label{final-comments}

In this paper we have proposed an index, named COVINDEX, that can be used for near real-time monitoring of COVID-19 pandemic. The index is computed as the ratio of the estimated test positive rate on a given day with respect to the value estimated for a week before.
Estimation of test positive rates is obtained by statistical modelling the daily empirical positive rates calculated from the observed data.
To this end, a GAM beta regression model with weights proportional to the administered tests is fitted.
By exploiting the relationship of penalized likelihood for GAMs with MAP Bayesian estimation, credible intervals for COVINDEX can be obtained via simulation to express the associated uncertainty.

We applied the proposed methodology to the Italian COVID-19 outbreak and we compared the trend of COVINDEX to the effective reproduction number $R_t$.
The analyses carried out confirm that $R_t$ is a delayed index of epidemic trend, and for this reason may provide a biased picture of the current pandemic status.
On the contrary, COVINDEX seems to provide a more up-to-date information which can be used as a decision-making tool. This aspect is of crucial importance for all policy makers and public health officials.
We defer to future research the evaluation of the implications deriving from the adoption of the proposed index.

Although the main focus of the analysis in this paper was the national level, similar considerations can be made for territorial administrative divisions, such as regions and provinces. In these cases, however, it should be noted that further assumptions are necessary, in particular the independence of the epidemic trend between neighbouring territories. 
However, an improved approach should account for the spatio-temporal dependency structure. For instance, \citet{Mingione:etal:2021} fitted a spatio-temporal CAR model with spatial dependence expressed by specifyicing an adjacency matrix derived from a network model of links and transport exchanges among Italian regions \citep{DellaRosa:etal:2020}. The study of these aspects is deferred to future research.

All the analyses have been performed in \texttt{R} version 4.1.0 \citep{Rstat}, using the package \texttt{mgcv} \citep{mgcv} and functions written by the author. Code to reproduce the analyses is available in a GitHub repository at \url{https://github.com/luca-scr/COVINDEX}.



\renewcommand\refname{References}
\bibliography{covindex.bib}

\end{document}